%% file: mdiqmv.tex
\documentclass[amsmath,amssymb,aps,twocolumn]{revtex4-2}

\usepackage{amssymb}
\usepackage{natbib}
\usepackage{graphicx}
\usepackage{amsmath}
\usepackage[bookmarks = false]{hyperref}
\usepackage{bm}
\usepackage[all]{hypcap}
\usepackage{graphicx}
\usepackage{colortbl}
\usepackage{booktabs}
\usepackage{enumerate}
\usepackage{enumitem}
\usepackage{braket}
\usepackage{mathrsfs}
\usepackage{multirow}
\usepackage{upgreek}
\usepackage{textcomp}
\newcommand{\degree}{^\circ}
\usepackage{soul}
\usepackage{dsfont}
\usepackage{tikz}

\usetikzlibrary{
  arrows.meta,
  calc,
  fit,
  positioning,
  quotes,
  shapes.geometric,
  shapes.multipart
}

\usepackage{array}
\newcommand{\PreserveBackslash}[1]{\let\temp=\\#1\let\\=\temp}
\newcolumntype{C}[1]{>{\PreserveBackslash\centering}p{#1}}
\newcolumntype{R}[1]{>{\PreserveBackslash\raggedleft}p{#1}}
\newcolumntype{L}[1]{>{\PreserveBackslash\raggedright}p{#1}}

\def\arraystretch{1.2}

\begin{document}

\title{Measurement-Device-Independent Verification of a Quantum Memory}

\author{Yong Yu$^{1,\,2}$}
\author{Peng-Fei Sun$^{1,\,2}$}
\author{Yu-Zhe Zhang$^{1,\,2}$}
\author{Bing Bai$^{1,\,2}$}
\author{Yu-Qiang Fang$^{1,\,2}$}
\author{Xi-Yu Luo$^{1,\,2}$}
\author{Zi-Ye An$^{1,\,2}$}
\author{Jun Li$^{1,\,2}$}
\author{Jun Zhang$^{1,\,2}$}
\author{Feihu Xu$^{1,\,2}$}
\author{Xiao-Hui Bao$^{1,\,2}$}
\author{Jian-Wei Pan$^{1,\,2}$}

\affiliation{$^1$Hefei National Laboratory for Physical Sciences at Microscale and Department
of Modern Physics, University of Science and Technology of China, Hefei,
Anhui 230026, China}
\affiliation{$^2$CAS Center for Excellence in Quantum Information and Quantum Physics, University of Science and Technology of China, Hefei, Anhui 230026, China}

\begin{abstract}
  In this paper we report an experiment that verifies an atomic-ensemble quantum memory via a measurement-device-independent scheme. A single photon generated via Rydberg blockade in one atomic ensemble is stored in another atomic ensemble via electromagnetically induced transparency. After storage for a long duration, this photon is retrieved and interfered with a second photon to perform joint Bell-state measurement~(BSM). Quantum state for each photon is chosen based on a quantum random number generator respectively in each run. By evaluating correlations between the random states and BSM results, we certify that our memory is genuinely entanglement-preserving.
\end{abstract}

\maketitle

Photonic quantum memory~\cite{lvovsky2009} is an enabling technique for quantum internet~\cite{wehner2018} and optical quantum computing~\cite{kok2007}, by providing capability of storing arbitrary single-photon states. Although basic memory operations have been realized in numerous experiments (e.g.~\cite{Liu2001a,Matsukevich2006,Choi2008,Hedges2010,Specht2011,Clausen2011,Zhang2011a,Saglamyurek2011b,ritter_elementary_2012,Sprague2014,Ding2015a,cho_highly_2016,yang_high_2015,yang_efficient_2016,wang_efficient_2019,zhong_nanophotonic_2017,maring_photonic_2017,pu_experimental_2017,bhaskar_experimental_2020,cao_efficient_2020,wang_cavity-enhanced_2021}), a convincing verification that can certify a quantum memory even from untrustful providers is still missing. In previous studies, it is typical of measuring amplitude decay of retrieval field, phase coherence in a superpositional basis, or performing quantum tomography to verify the storage capability. However, such measurements may be cheated simply if the memory supplier is untrustful. For instance, a cheater can counterfeit a quantum memory with unitary storage fidelity and infinite long lifetime merely by using an intercept-and-resend scheme as long as the measurement basis is known to the cheater in advance~(see Supplemental Material). It is thus necessary of making the input state and output measurement basis randomly selected. Even in this situation, the counterfeit scheme can still work by compromising a lower memory efficiency, by taking advantage of imperfect devices in the measurement setup, similar to situations in the field of secure quantum key distribution (QKD)~\cite{xu_secure_2020}.

It is ideal of using the Bell test~\cite{brunner2014} to verify a quantum memory by storing one photon from an entanglement pair. Violation of Bell inequality certifies that the memory is entanglement preserving. This scheme relies solely upon statistical correlations and puts no assumption on the apparatus. Therefore it is immune to any attack and called as device-independent (DI). Nevertheless, it is technically very high demanding to implement the Bell scheme, since it is required to close the loopholes~\cite{hensen2015}. First, the two measurement apparatus needs to be space-like separated, which is a too strong requirement for long-lived quantum memories. Second, the overall efficiency (including storage, transmission, and detection efficiency) needs to be higher than a threshold value (e.g.~$82.8\%$ in \cite{larsson1998}) to close the fair sampling loophole, which excludes plenty of memories with moderate and low efficiency. So far, all the reported experiments of entanglement storage~(e.g.~\cite{Matsukevich2006,Zhang2011a,Clausen2011,Saglamyurek2011b,ritter_elementary_2012,Ding2015a,cao_efficient_2020}) are still very far away from reaching this criteria.

\begin{figure*}[htbp]
	\centering
	\includegraphics[width=0.9\textwidth]{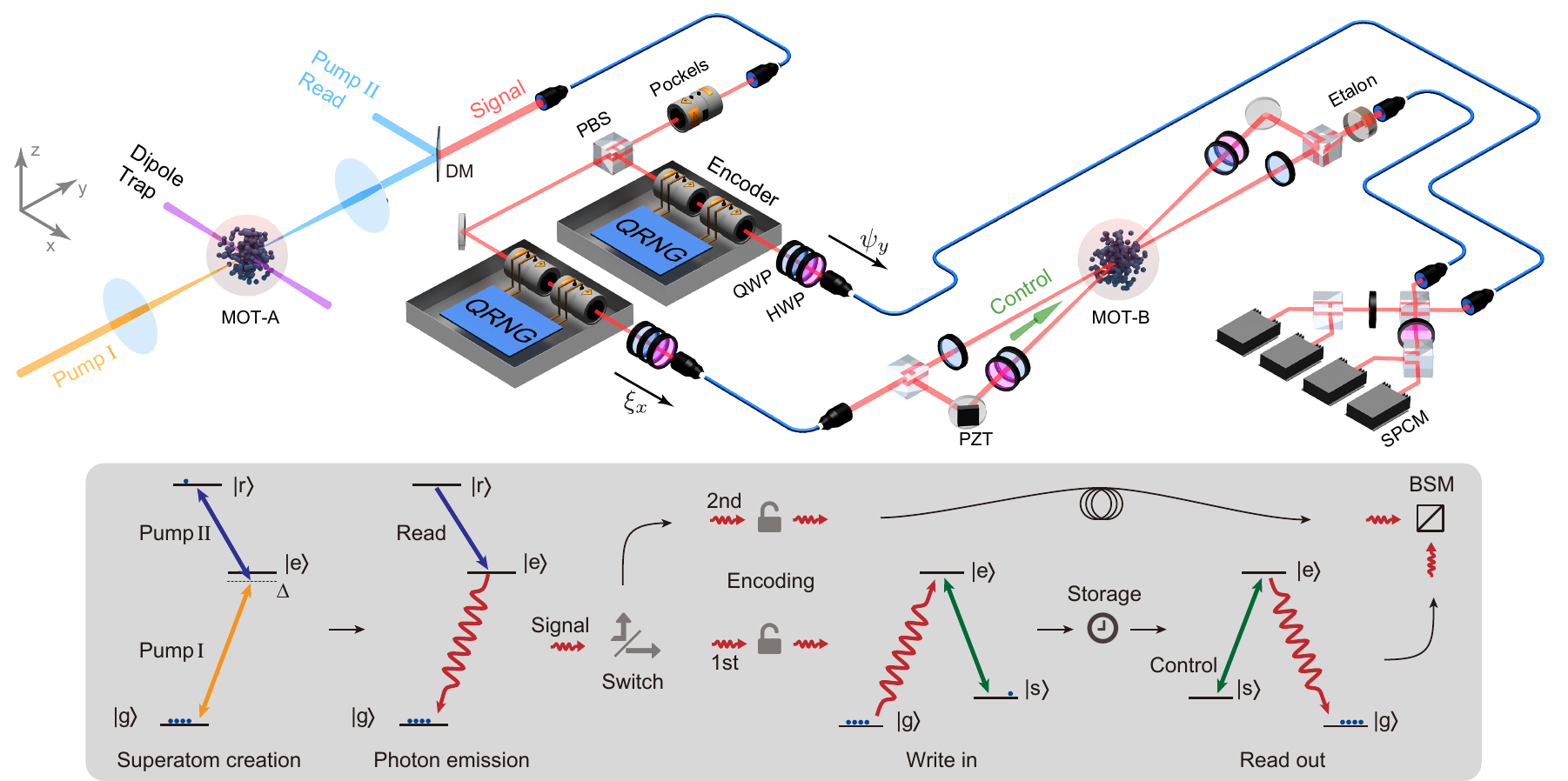}
	 \caption{Experimental setup and energy level scheme (gray shaded area). MOT-A and B are two laser-cooled $^{87}$Rb atomic ensembles, serving as the single-photon source and the quantum memory, respectively. MOT-A is further loaded into a small dipole trap to confine its dimension. For both MOT-A and B, a slight bias magnetic field is applied along y direction to remove the degeneracy of Zeeman sublevels, and all atoms are prepared in the ground state $\ket{g}$ initially. In MOT-A, only one atom is excited to Rydberg state $\ket{r}$ by beam Pump \uppercase\expandafter{\romannumeral1} and \uppercase\expandafter{\romannumeral2}, forming a Rydberg superatom. It is soon retrieved into a signal photon by the Read beam. The photon is dynamically switched to two paths with/without the quantum memory via a Pockels cell. In each path, an encoder consisted of two Pockels cells controlled by a QRNG encodes a polarization qubit onto this photon. The first photon is sent into MOT-B to store. This is achieved by mapping it onto a spin-wave on $\ket{s}$ with the help of the Control beam. After a duration of storage, it is retrieved and interferes with the second photon in the BSM device. In MOT-B, two polarization modes of a photon are mapped onto two space modes before storing and recombined after retrieving. Two space modes are actively phase-locked by adjusting the PZT. DM: dichroic mirror. PBS: polarizing beamsplitter. QWP: quarter-wave plate. HWP: half-wave plate. SPCM: single-photon counting module.}
	\label{fig:setup}
\end{figure*}

One may relax the technical overheads, by trusting some part of the verification components. A similar idea was implemented in the field of QKD by designing a measurement-device-independent (MDI) scheme~\cite{lo2012,Braunstein2012}. It excludes most common attacks towards measurement setup, meanwhile keeps a moderate level of experimental feasibility. The MDI-QKD has been very successful and improves practical QKD security significantly. Furthermore, the MDI scheme has successfully advanced the field of entanglement witness~\cite{branciard2013,Verbanis2016,li2020}. In the case of memory verification, it is also very natural of trusting the state preparation process. Rosset \textit{et al} laid out a theoretical framework on how to verify a quantum memory via MDI~\cite{rosset2018}. Two experiments~\cite{mao2019,Graffitti2020} in previous have been carried out to verify a quantum channel of optical fiber. Here, we perform the first MDI verification of a genuine quantum memory.

In the MDI scheme, Alice, the client, verifies a quantum memory $\mathcal{N}$ afforded by an untrustworthy supplier, Bob, by playing a semi-quantum signaling game with him. In each round, Alice sends two quantum questions $\xi_x$ and $\psi_y$, according to the random number $x$ and $y$, sequentially to Bob and ask him for a classical answer $b$. In the language of quantum optics, $\xi_x$ and $\psi_y$ are two photonic qubits. Bob performs a joint BSM towards them and gives the result $b$. Preparing a photon in $\xi_x$ ($\psi_y$) could be viewed as a virtual process that preparing an entanglement between this photon and a virtual photon in state $\ket{\Phi^+}=(\ket{00}+\ket{11})/\sqrt{2}$, then projecting the virtual photon onto $\xi_x$ ($\psi_y$). The BSM could, therefore, be viewed as an entanglement swapping. It makes two virtual photons onto an entangled state $\ket{\Phi^+}\bra{\Phi^+}$. With one photon under the storage of the memory, the entangled state between two virtual photons become $\mathbf{J}_{\mathcal{N}}=(\mathcal{N}\otimes\openone)\ket{\Phi^+}\bra{\Phi^+}$, which is exactly the Choi matrix of the memory~\cite{jiang2013}. Hence we can verify a non-EB memory when witness two virtual photons entangled, i.e. $\Braket{W}=Tr[\mathbf{J}_{\mathcal{N}}W]>0$, where $W$ is the entanglement witness operator~\cite{guhne2009}.

\begin{figure*}[hbt]
	\centering
	\includegraphics[width=0.7\textwidth]{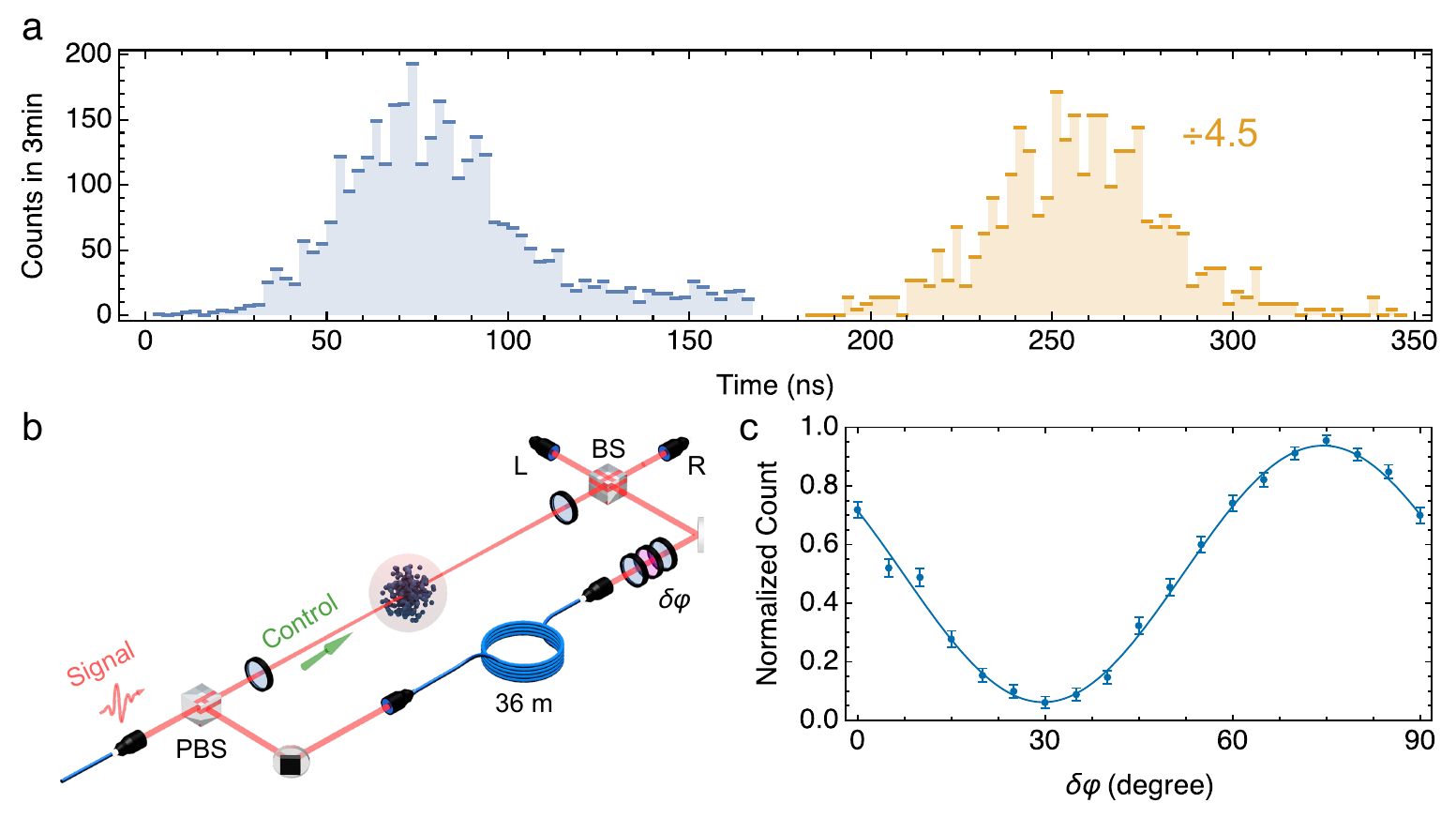}
	\caption{Characterization of photon distortion during storage. a. The temporal waveform of the photon before (blue) and after (orange) storage. The counts of post-storage are amplified by a factor of $4.5$. b. A Mach-Zehnder interferometer to detect homogeneity between stored and unstored photons. One arm is the EIT quantum memory, and the other arm is a 36~m fiber delay line. In the delay line arm, an HWP surrounded by two QWPs is inserted. The relative phase between two arms will vary along with the angle $\delta\varphi$ of the HWP. c. The oscillation of counts in port L along with the angle $\delta\varphi$ of the HWP.
	 }
	\label{fig:shape}
\end{figure*}

The setup of our experiment is shown in Fig.~\ref{fig:setup}. The quantum memory we verified is based on the two-channel electromagnetically induced transparent (EIT) mechanism~\cite{fleischhauer2005} in a laser-cooled $^{87}$Rb atomic ensemble. In the beginning, all atoms are prepared on the ground state $\ket{g}\equiv\ket{5S_{1/2},F=2,m_F=+2}$, which is not degenerate with other sublevels due to a slight bias magnetic field. A signal photon, with a qubit encoded on its polarization, is spatially split by a polarizing beamsplitter (PBS). In each spatial mode, polarization is rotated to $\sigma^+$ for atoms to couple transition $\ket{g}\leftrightarrow\ket{e}\equiv\ket{5P_{1/2},F=1,m_F=+1}$. Each signal beam and the control beam have the waist 90~$\upmu$m and 300~$\upmu$m, overlapping well in the atomic region with 3$^{\degree}$ angle. In storing phase, the control beam coupling $\ket{e}\leftrightarrow\ket{s}$ dynamically converts the signal photon into a spin-wave on $\ket{s}$ and is switched off. After a duration of storage, two spatial modes of the signal photon are simultaneously retrieved by turning the control beam on, and combined to a polarization encoded one again. For two spatial modes, storage efficiency $\eta_s=0.269$ and $0.285$, and memory lifetime $\tau_m=58.2$~$\upmu$s and 56.6~$\upmu$s, showing a good conformity between them (see Supplemental Material). The relative phase shift between two spatial modes is actively stabilized to avoid dephasing. To depress the noise, the retrieved photon is filtered via an etalon with linewidth $\sim0.5$~GHz.

On a time scale of dozens of $\upmu$s that we care about, the phase walk-off between two spatial modes of the memory is neglectable small. The more dominant disturbance is the depolarizing noise, including residual control photons and dark counts of the detectors. These noises, notwithstanding small enough, dominate when signal dissipating along with storing. Thus we model our memory as a depolarizing channel as
\begin{equation}\label{eq:channel}
	\mathcal{N}(\rho)=(1-p)\rho+p\frac{\mathds{1}}{2},
\end{equation}
where $\openone$ is the identity operator. Its Choi matrix is $\mathbf{J}_{\mathcal{N}}=(1-p)\left|\Phi^{+}\right\rangle\left\langle\Phi^{+}\right|+p(\mathds{1}$/4). Based on these, the detailed experimental scheme is designed as following. State $\xi_x$ ($\psi_y$) is randomly picked up from a set $\{\ket{H},\ket{V},\ket{D},\ket{R}\}$, according to a random number $x$ ($y$). $\ket{H}$ and $\ket{V}$ stand for horizontal and vertical polarization, respectively. $\ket{D}\equiv(\ket{H}+\ket{V})/\sqrt{2}$ and $\ket{R}\equiv(\ket{H}+i\ket{V})/\sqrt{2}$ are $+45^{\circ}$ linear polarization and right circular polarization, respectively. The BSM of two photonic qubits is achieved by overlapping them on a PBS and performing X basis measurement at each output port. It could differentiate $\ket{\Phi^{\pm}}=(\ket{HH}\pm\ket{VV})/\sqrt{2}$ from all two qubit states. Let $b=+$ and $-$ for this two results. We can derive their payoff function as
\begin{equation}
	\begin{gathered}
	w^{+}_{xy}=
	\begin{pmatrix}
	0 & -\frac{1}{2} & -\frac{1}{2}&\frac{1}{2} \\ -\frac{1}{2}& 0 & -\frac{1}{2}&\frac{1}{2}\\
	 -\frac{1}{2} & -\frac{1}{2}& 1 &0\\
	\frac{1}{2} & \frac{1}{2}&0&-1 \end{pmatrix},\:
	w^{-}_{xy}=
	\begin{pmatrix}
	 0 & -\frac{1}{2} & \frac{1}{2}&-\frac{1}{2} \\
	 -\frac{1}{2}& 0 & \frac{1}{2}&-\frac{1}{2}\\
	\frac{1}{2} & \frac{1}{2}& -1 &0\\
	-\frac{1}{2} & -\frac{1}{2}&0&1 \end{pmatrix}
	\end{gathered},
\end{equation}
by decomposing corresponding witness operators~\cite{rosset2018}. For other results not interested, let $b=0$ and $w^0=0$. Through repetition, Alice can do statistic of probability of each combination $P(b|xy)$ and calculate average payoff $\Braket{W}=\sum_{xy}P(b|xy)\omega^b_{xy}$.

The quality of our BSM depends on the homogeneity between two input photons, i.e. a stored and an unstored one in our case. In EIT quantum memory, the temporal waveform of a photon stored will be largely affected by the envelope of the control beam. With commonly used step function envelope of the control beam, a Gaussian waveform input photon after storage will be distorted\cite{gorshkov2007}. To eliminate the distortion, we use an optimized envelope of the control beam by following the approach proposed by Gorshkov \textit{et al}.~\cite{gorshkov2007,gorshkov2008} The envelope is manipulated by dynamical modulation of the radio frequency signal driving the acoustic optical modulator (AOM). Fig.~\ref{fig:shape}.~a shows good waveform conformity of the photon before and after storage. We further check the homogeneity between them via a Mach-Zehnder interferometer depicted in Fig.~\ref{fig:shape}~b. Two arms of this interferometer are the EIT quantum memory and a 36~m fiber delay line. A single photon will travel both paths simultaneously and interfere with itself on a beamsplitter (BS). In the fiber delay arm, we insert an HWP as well as two QWPs surrounded. The relative phase between two arms will change along with the angle $\delta\varphi$ of the HWP. Thus we can observe an oscillation of counts in port L and R. From the oscillation in port L shown in Fig.~\ref{fig:shape}.~c, the interference visibility is deduct as $V=0.875\pm 0.018$. We can model the influence of inhomogeneity as a noise in BSM and write the realistic BSM operator as $S^{\pm}=(1-\lambda)\ket{\Phi^{\pm}}\bra{\Phi^{\pm}}+\lambda\ket{\Phi^{\mp}}\bra{\Phi^{\mp}}$, where parameter $\lambda=1-V^2$ (see Supplemental Material).

\begin{figure}[tbp]
	\centering
	\includegraphics[width=0.75\columnwidth]{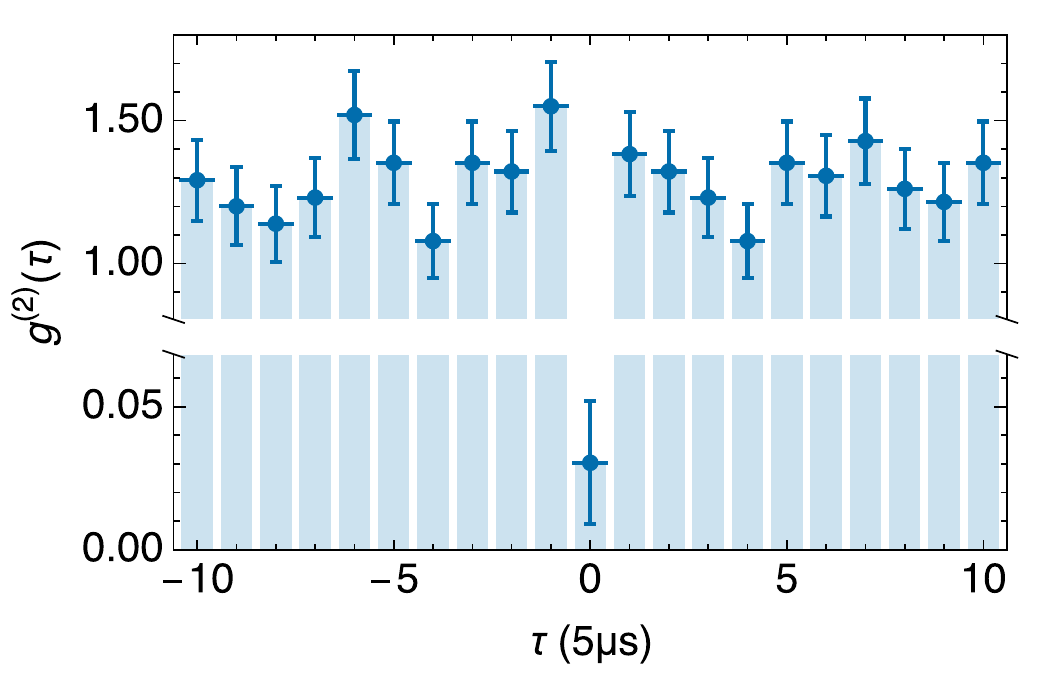}
	\caption{Second-order correlation function $g^{(2)}$ of the Rydberg single-photon source. Results are measured by a Hanbury–Brown-Twiss experiment. The error bars represent one standard deviation.}
	\label{fig:hbt}
\end{figure}

Although measurement devices are unconditionally safe in the MDI scheme, we still need to secure a true single-photon source in case of the photon number splitting attack~\cite{brassard2000}. To achieve this, we generate single photons in a small ensemble of $^{87}$Rb with the help of the Rydberg blockade~\cite{saffman2010,Dudin2012}. All atoms, loaded in a small volume dipole trap, are initially prepared in the ground state $\ket{g}$. Two beams, Pump \uppercase\expandafter{\romannumeral1} and \uppercase\expandafter{\romannumeral2}, coupling $\ket{g}\leftrightarrow\ket{e}$ and $\ket{e}\leftrightarrow\ket{r}$ respectively (with single-photon detuning $\Delta=40$~MHz), transfer atoms to a highly excited Rydberg state $\ket{r}\equiv\ket{81S_{1/2},m_j=1/2}$ via a two-photon Raman process. Thanks to the strong blockade effect, only one atom within the ensemble could be excited, forming a Rydberg superatom~\cite{Dudin2012}, $\frac{1}{\sqrt{N}}\sum_i\ket{g_1\dots r_i\dots g_N}$. The superatom is then converted to a signal photon under the irradiation of a read beam coupling $\ket{r}\leftrightarrow\ket{e}$. Different from the single atom, the superatom will emit photon to a predefined phase-matching spatial mode with no necessity of an optical resonator. In our setup, the single-photon efficiency is $\sim6\%$ (defined as the probability to have a photon before the path switching Pockels cell in each trial). Through a Hanbury-Brown-Twiss experiment, we know the second-order correlation of the signal photon is lower to $g^{(2)}(0)=0.03\pm0.02$ as shown in Fig.~\ref{fig:hbt}, demonstrating a good nonclassical antibunching property.

To further strengthen the immunity to attacks, we perform a fully random qubit preparation. In each round, we generate two photons chronologically. Two photons are dynamically switched to different paths by a Pocekls cell and a PBS. In each path, two Pockels cells, serving as dynamical HWP and QWP respectively, are used to encode a polarization qubit $\xi_x$ ($\psi_y$). Their status is controlled by the random number $x$ ($y$). We pick up $x$ and $y$ right before they are used from a quantum random number generator (QRNG) with 10~MHz repetition rate (see ref.~\cite{nie2015} for details).

\begin{figure}[tp]
	\centering
	\includegraphics[width=0.75\columnwidth]{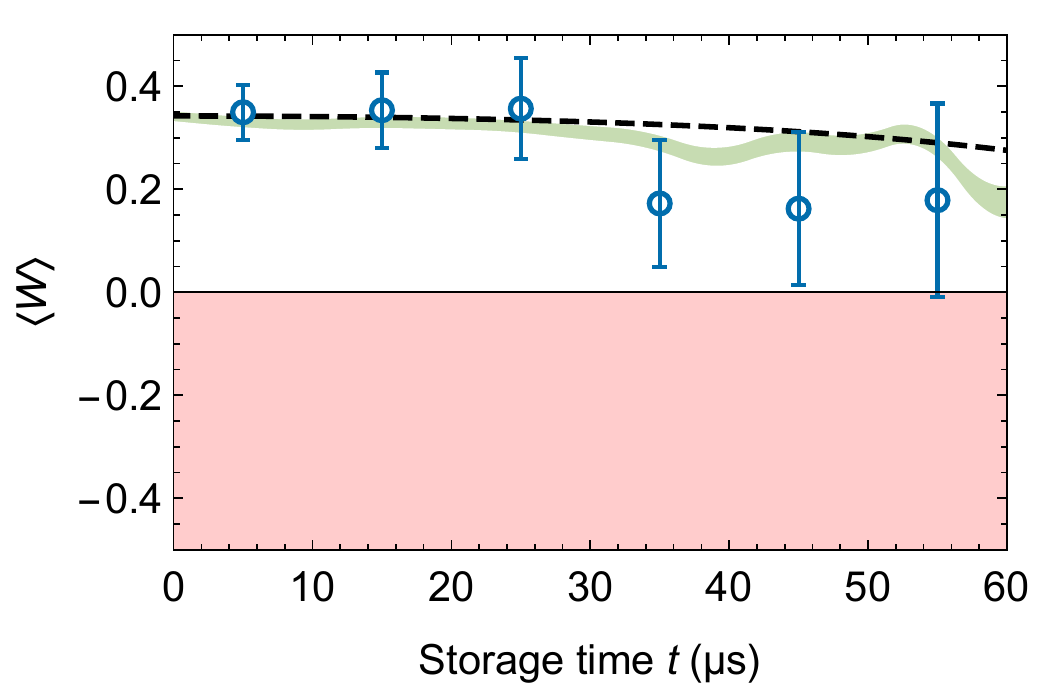}
	\caption{MDI verification results of the quantum memory. The blue circles are the tested results. The error bars represent one standard deviation. The black dash line is the theoretical predictions. The green shadow is the one standard deviation region of the simulated results. Pink shading indicates the entanglement-breaking regime.
	}
	\label{fig:payoff}
\end{figure}

Blues circles in Fig.~\ref{fig:payoff} shows the MDI verification results of the quantum memory. To get rid of the influence of optical and detection losses, we normalize $P(b|xy)$ by only considering detected events. Within the memory lifetime, we always have $\Braket{W}>0$ to witness a quantum memory. Along with storage time $t$ increasing, $\Braket{W}$ decreases because of the depolarizing noise. Meanwhile, we get fewer data for two reasons. One is that the retrieval efficiency of the quantum memory gradually drops. The other is that the period time of one cycle increases, leading to a lower repetition rate. Hence the uncertainty of results increases in long storage time cases.

To verify the validation of the MDI results, we compare them with the theoretical predictions and the simulated results (see Supplemental Material). First, we measure the parameters of our system, including the efficiency and lifetime of the quantum memory, background noise, as well as optical efficiency. Based on these, we can predict the noise strength $p$ in Eq.~\ref{eq:channel} as a function of $t$, and further calculate $\Braket{W}$. This method takes the memory as a pure depolarizing channel and set the upper bound of the tested results. Second, we directly characterize the quantum memory via process tomography in different $t$. With the knowledge of the process matrices $\chi$, we can simulate the results. The theoretical and simulated results are shown as the black dash line and green shadow in Fig.~\ref{fig:payoff}, respectively. We can see that when $t<30$~$\upmu$s, the measured results are in good agreement with the theoretical predictions and the simulated results. When $t>30$~$\upmu$s, the measured and simulated results are getting lower than the theoretical predictions, indicating that other noise possibly gradually starts to dominate. The measured and simulated results are consistent with the margin of error throughout the measurement interval.

In summary, we have performed the first verification of a quantum memory via the MDI scheme. Reliability is enforced by making use of a single-photon source and fully random state preparation. The results show good consistency with the theoretical predictions. Similar methods can be straightforwardly applied to other kinds of quantum memories and quantum components, such as quantum frequency converters, quantum nondemolition measurement devices. One drawback in the current experiment is that the single-photon efficiency is rather low, leading to a slow verification process and large measurement uncertainties. By using state-of-the-art single-photon source~\cite{wang2019Towards} with the efficiency of $\sim60\%$, the MDI approach will be as efficient as quantum process tomography. With this improvement, the MDI approach may serve as a standard verification of the quantum memories or other devices in large-scale quantum internet in the future.

This work was supported by National Key R\&D Program of China (No.~2017YFA0303902, No.~2020YFA\-0309804), Anhui Initiative in Quantum Information Technologies, National Natural Science Foundation of China, and the Chinese Academy of Sciences.

\clearpage

\setcounter{figure}{0}
\setcounter{table}{0}
\setcounter{equation}{0}

\onecolumngrid

\global\long\def\theequation{S\arabic{equation}}
\global\long\def\thefigure{S\arabic{figure}}
\renewcommand{\thetable}{S\arabic{table}}

\newcommand{\msection}[1]{\vspace{\baselineskip}{\centering \textbf{#1}\\}\vspace{0.5\baselineskip}}

\msection{SUPPLEMENTAL MATERIAL}

\section{An attack scheme to process tomography}
\input{attack.tex}
By using the faked-state attack~\cite{makarov2011}, we design a scheme to fake a quantum memory that could pass the process tomography verification as shown in Fig.~\ref{fig:attack}. There are four steps in standard process tomography~\cite{nielsen2002}. First, Alice prepare a single photon belong to a tomographical complete set, e.g. $\{\ket{H},\ket{V},\ket{D},\ket{R}\}$ and send it to the ``quantum memory". Second, she sends the qubit into the memory to store. After qubit retrieved, third, she chooses a basis from $X$, $Y$, and $Z$ and fourth, perform the measurement. Inside the ``quantum memory'', Bob directly perform the measurement in a randomly chosen basis right after the qubit is input, and get the result $R$. When Alice chooses a measurement basis, Bob eavesdrops this message and compares it with the basis he chooses. If their base are same, Bob manipulates Alice's detector to output a counterfeit result $R$. Otherwise, he makes the detector output nothing, pretending a failed storage.

For the same base case, it is equivalent for Alice to do a tomography to unprocessed qubits. She will always conclude an identity process, i.e. a quantum memory with unitary fidelity and infinite long lifetime. If Alice chooses measurement in a fully random fashion, Bob has a probability of $1/3$ to guess the same basis with her, faking a memory with the corresponding efficiency. If Alice chooses the basis in not random or in a passive fashion, Bob could know her basis before his measurement. Thus he can always give the output, making the memory efficiency up to $100\%$.

\section{Technical details}

MOT-A is first cooled and collected by a conventional magneto-optical trap for 30~ms. In order to increase the density, it is then switched to the `dark SPOT' configuration~\cite{ketterle1993} in the next 70~ms, by employing counterpropagating hollow repumper beams with $4\times4$~mm holes in two orthogonal directions and a depumper beam. The depumper beam is coupling $\ket{5S_{1/2},F=2}\leftrightarrow\ket{5S_{3/2},F=2}$ transition, whereas repumper is tuned from $\ket{5S_{1/2},F=1}\leftrightarrow\ket{5S_{3/2},F=2}$ to $\ket{5S_{1/2},F=1}\leftrightarrow\ket{5S_{3/2},F=1}$ transition. At the last 3~ms of `dark SPOT' cooling, the intensity of the repumper beam is gradually reduced to benefit from the temporal `dark SPOT' mechanism. After this stage, we get a cloud of atoms with $1/e^2$ waist of 378~$\upmu$m and density of $9.3\times 10^{10}$~cm$^{-3}$. Atoms are next loaded to an optical dipole trap along x-direction employing 1064~nm laser with the intensity of 2~W and the waist of 25~$\upmu$m (y-direction) and 5~$\upmu$m (z-direction). With the help of 22~ms of trapping, the ensemble radius is confined to 27.2~$\upmu$m in the y-direction, with a density of $4.13\times10^{11}$~cm$^{-3}$. After 122~ms of cooling and trapping, trails of single-photon generation repeat at the frequency up to 200~kHz in the following 3~ms until the next cooling phase. The 475~nm beams (Pump \uppercase\expandafter{\romannumeral1} and Read) and 795~nm beam (Pump \uppercase\expandafter{\romannumeral2}) have a waist of 7.7~$\upmu$m and 7~$\upmu$m, respectively. They propagate in the opposite direction and overlap with each other through the atoms. MOT-B is synchronously cooled with MOT-A. With 122~ms conventional magnetic-optical trap and temporal `dark SPOT' in the last 3~ms, the optical depth of two signal modes is estimated as 9.75 and 9.62, respectively.

\section{Imperfection of Bell state measurement}
\begin{figure}[htbp]
    \centering
    \includegraphics[width=0.9\textwidth]{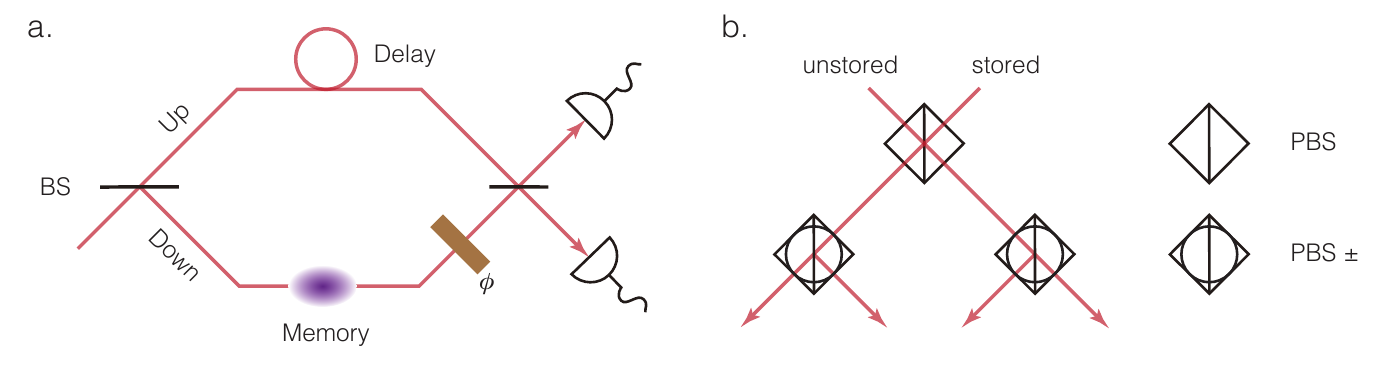}
    \caption{a. The Mach-Zenhder interferometer to detect photon homogeneity. b. The BSM setup.}
    \label{fig:bsm}
\end{figure}

We use a Mach-Zehnder interferometer as shown in Fig.~\ref{fig:bsm}.~a to detect the homogeneity between stored and unstored photons. After the first beamsplitter, we have
\begin{equation}
    a_D^{\dagger} \to a_U^{\dagger} + ia_D^{\dagger}.
\end{equation}
$a^{\dagger}$ is the creation operator of a Gaussian shape photon. Subscript $U$ and $D$ indicate up and down space mode. We model the distortion of photon during storage by replacing creation operator $a^{\dagger}$ with $b^{\dagger}=\alpha a^{\dagger} + \beta \tilde{a}^{\dagger}$, where $\tilde{a}^{\dagger}$ is a creation operator in the orthogonal space of $a^{\dagger}$. We can calculate the state after the second beamsplitter as
\begin{equation}
    \begin{split}
        & a_U^{\dagger} + ib_D^{\dagger} \\
        \to & \left(a_D^{\dagger} + ia_U^{\dagger}\right) + ie^{i\phi}\alpha\left(  a_U^{\dagger} + i a_D^{\dagger} \right) +ie^{i\phi}\beta\left(\, \tilde{a}_U^{\dagger} + i \tilde{a}_D^{\dagger} \right)\\
        =&(1 - \alpha e^{i\varphi})a_D^{\dagger} + i(1 + \alpha e^{i\varphi})a_U^{\dagger} - e^{i\phi}\beta \tilde{a}_D^{\dagger} + ie^{i\phi}\beta \tilde{a}_U^{\dagger}.
    \end{split}
\end{equation}
By varying the phase $\phi$, we can observe the oscillation of counts in two ports. The oscillation visibility is given by
\begin{equation}
    V = \frac{N_{\max } - N_{\min } }{N_{\max } - N_{\min }} = \frac{[(1 + \alpha)^2 + \beta^2] - [(1 - \alpha)^2 + \beta^2]}{[(1 + \alpha)^2 + \beta^2] +[(1 - \alpha)^2 + \beta^2]} = \alpha.
\end{equation}
$N$ with subscript max (min) represents maximal (minimal) counts of the oscillation pattern.

We then consider the influence of inhomogeneity upon BSM. Our BSM apparatus is shown in Fig.~\ref{fig:bsm}.~b. The first PBS is to post-select two photons both in horizontal or vertical polarization. Two PBS in $\ket{+/-}=\ket{H}\pm \ket{V}$ basis discriminate $\ket{\Phi^+}$ from $\ket{\Phi^-}$. Considering an imperfect interference, we may misrecognize two Bell states. Thus we can model a realistic BSM as
\begin{equation}
    S^{\pm} =(1 - \lambda)\ket{\Phi^{\pm}}\bra{\Phi^{\mp}} + \lambda\ket{\Phi^{\mp}}\bra{\Phi^{\mp}}.
\end{equation}
We write these two bell states by creation operators as
\begin{equation}
    \begin{split}
    \ket{\Phi^{\pm}}&=a^{\dagger}_Hb^{\dagger}_H\pm a^{\dagger}_Vb^{\dagger}_V\ket{vac}\\
    &=\left(\alpha(a^{\dagger}_Ha^{\dagger}_H+a^{\dagger}_Va^{\dagger}_V)\pm\beta(a^{\dagger}_H\tilde{a}^{\dagger}_H+a^{\dagger}_V\tilde{a}^{\dagger}_V)\right)\ket{vac},
    \end{split}
\end{equation}
where the subscript represents the polarization of the photon. The first term corresponds to two homogeneous photons interfering on $\ket{+/-}$ basis PBSs, i,e, an ideal BSM. The second term corresponds to two distinguishable photons interfering on $\ket{+/-}$ basis PBSs. This would lead to a random coincidence, i.e. Bell states being misrecognized by $1/2$ probability. Therefore we get $\lambda=\beta^2/2=(1-V^2)/2$.

\section{Result analysis}
\subsection{Theoretical result}
\begin{figure}[htbp]
    \centering
    \includegraphics[width=0.6\textwidth]{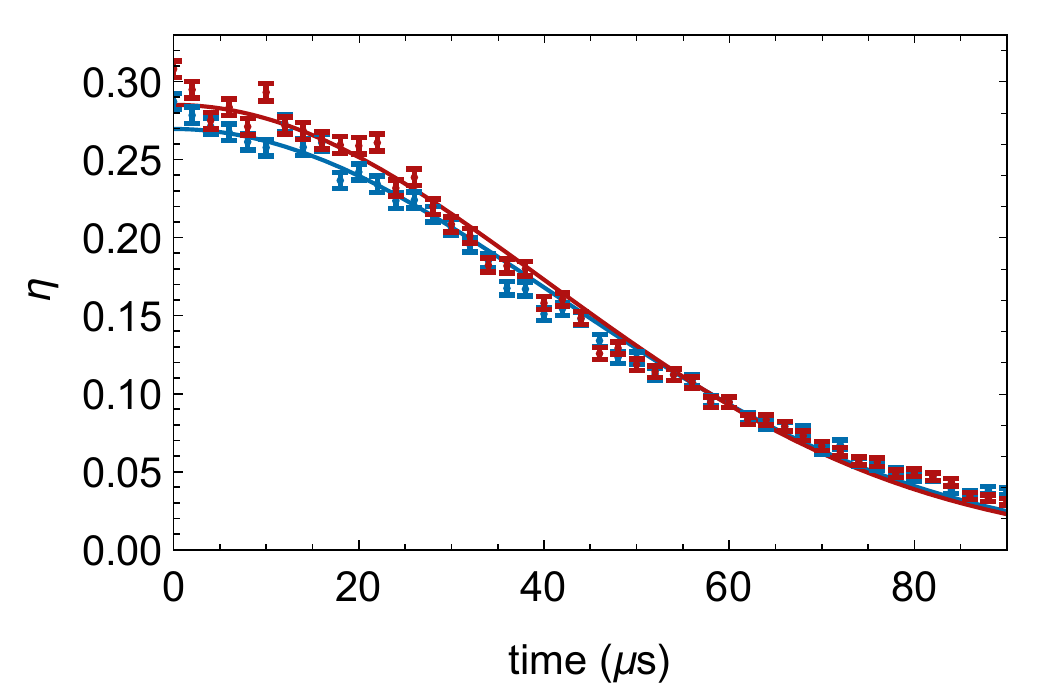}
    \caption{Storage efficiency of M1 (red) and M2 (blue) spatial mode as a function of storage time $t$. The error bars represent one standard deviation. The curve with corresponding color is the exponential fitting.}
    \label{fig:lifetime}
\end{figure}

The noise strength $p$ of a pure depolarizing channel is closely related with the signal to noise ratio (SNR) as
\begin{equation}
    \frac{1 - p}{ p} = \text{SNR}.
\end{equation}
The observed signal probability is given by
\begin{equation}
    P_{signal} = P_{ph}\eta_{m}\eta_{opt}\eta_{det}.
\end{equation}
$P_{ph}$ is the single-photon efficiency (defined as the probability of having a photon before the path switching Pockels cell). $\eta_{m}$ is the storage efficiency of the EIT quantum memory. We label two spatial modes as M1 (transmission at the entrance PBS) and M2 (reflection at the entrance PBS) and measure their $\eta_{m}$ as a function of storage time $t$ as shown in Fig.~\ref{fig:lifetime}. By fitting the results with a exponential function $\eta_{m}=\eta_m^0 e^{-t^2/\tau_m^2}$, we can estimate the storage lifetime $\tau_{m}$ of each mode. $\eta_{opt}$ is the optical efficiency, including the efficiency of polarization encoding and photon transmission. $\eta_{det}$ is the detection efficiency of SPCMs. We also measure the probability of noise $P_{noise}$ by blocking the single-photon source. The tested value of these parameters are listed in Tab.~\ref{tab:results}.

\begin{table}[htbp]
    \renewcommand\arraystretch{0.6}
    \label{tab:results}
    \centering
    \begin{tabular}{C{2cm} C{2cm} C{2cm} C{2cm} C{2cm} C{3cm}}
    	\toprule
        $Pph$ & $\eta_{opt}$ & $\eta_{det}$ & $\eta^0_m$ & $\tau_m$($\upmu$s) & $P_{noise}$ \\
    	\midrule
        \multirow{2}{*}{$0.060$} & \multirow{2}{*}{$0.108$} & \multirow{2}{*}{$0.70$} & $0.269$ & $58.2$ & \multirow{2}{*}{$8.57\times10^{-5}$} \\
    	 &  &  & $0.250$ & $56.6$ &  \\
    	\bottomrule
    \end{tabular}
    \caption{Experimental parameters. $\eta_m^0$ and $\tau_m$ are separately listed for M1 (up) and M2 (down).}
\end{table}

In the MDI scheme, two photons are projected into an entangled state by the BSM. Therefore the postselected entangled state could be expressed by the projection operator, $\rho_{ent}=S^{+}$ for instance. Supposing one photon goes through a depolarizing channel and the other keeps unchanged. So the entangled state becomes
\begin{equation}
    \rho'_{ent} = (1 - p) \rho_{ent} + p \frac{\openone}{4}.
\end{equation}
Then we can calculate the performance for this depolarizing channel as
\begin{equation}
    \Braket{W} = \text{Tr}[W\rho'_{ent}].
\end{equation}

\subsection{Simulated result}
\begin{figure}[htbp]
    \centering
    \includegraphics[width=0.8\textwidth]{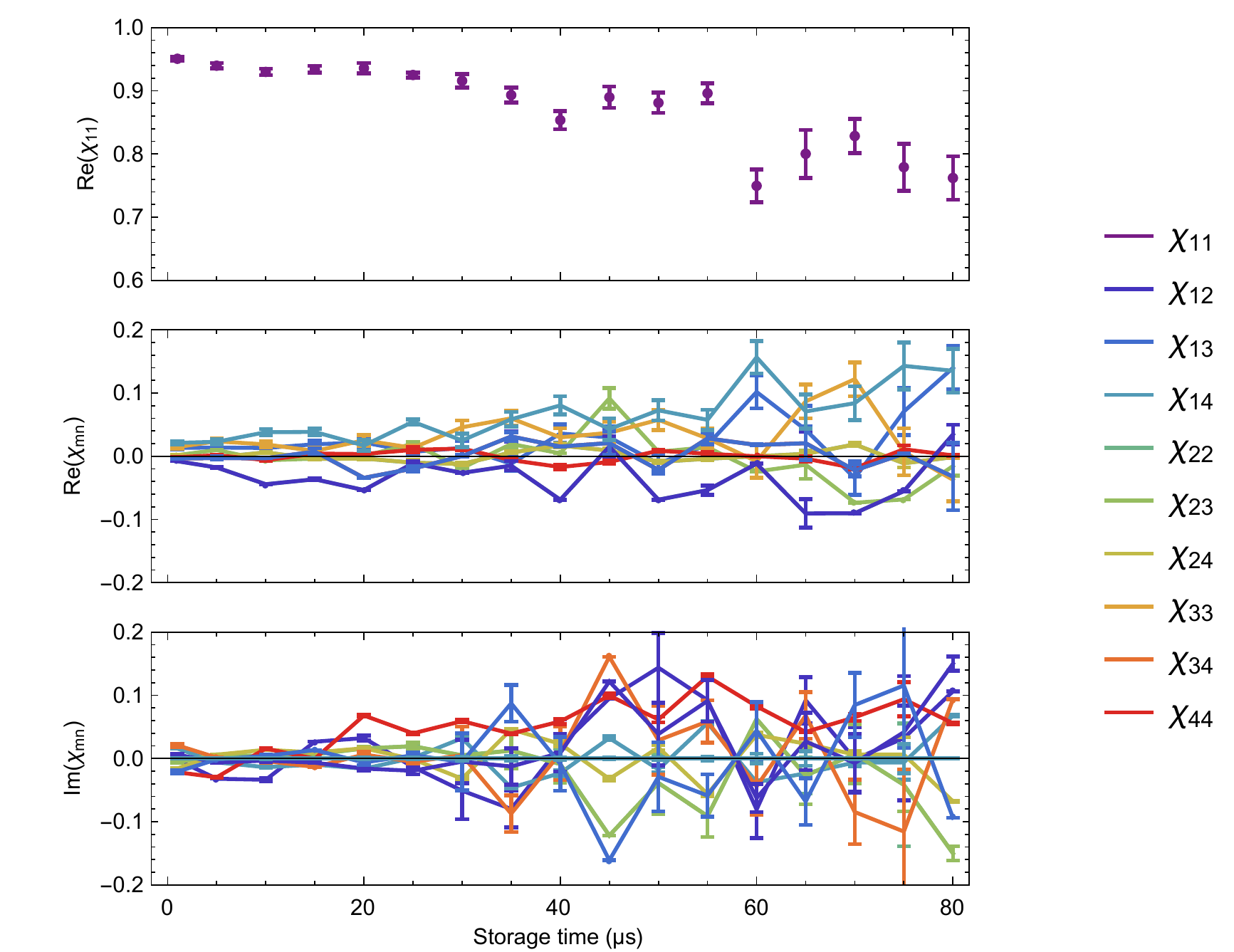}
    \caption{Results of $\chi_{mn}$.}
    \label{fig:chi}
\end{figure}

Similarly, we give the simulation by replacing the pure depolarizing channel model by the directly characterized process matrix. Generally, a single qubit process could be expressed as
\begin{equation}
    \mathcal{E}(\rho) =\sum_{mn}\tilde{E}_m \rho \tilde{E}_n^{\dagger}\chi_{mn}.
\end{equation}
When $\tilde{E}_0=\openone$, $\tilde{E}_1=\sigma_x$, $\tilde{E}_2=-i\sigma_y$, $\tilde{E}_3=\sigma_z$ are chosen, idealy, we should have $\chi_{11}=1$ and $\chi_{mn}=0$ when $m,n\neq1$ for a quantum memory. Fig.~\ref{fig:chi} shows the real $\chi_{mn}$ in our experiment by process tomography. Thus the $\rho_{ent}$ after storage could be expressed as
\begin{equation}
    \rho''_{ent} =\sum_{mn}(\tilde{E}_m \cdot \openone) \rho_{ent} (\tilde{E}_n \cdot \openone)^{\dagger}\chi_{mn}.
\end{equation}
Based on this, we can simulate the average witness as
\begin{equation}
    \Braket{W} = \text{Tr}[W\rho''_{ent}].
\end{equation}

\end{document}

%% file: attack.tex
\tikzset{
  meta box/.style={
    draw,
    black,
    very thick,
    text centered
  },
  punkt/.style={
    meta box,
    rectangle,
    rounded corners,
    inner sep=7pt,
    minimum height=2em,
    minimum width=5em,
    align=center,
    text width=5em
  },
  stepbox/.style={
    meta box,
    rectangle,
    inner sep=7pt,
    minimum height=2em,
    minimum width=8em,
    align=center,
    text width=9em
  },
  basisbox/.style={
    meta box,
    rectangle,
    inner sep=7pt,
    minimum height=2em,
    minimum width=5em,
    align=center,
    text width=6em
  },
  outbox/.style={
    meta box,
    rectangle,
    inner sep=2pt,
    minimum height=2em,
    minimum width=5em,
    align=center,
    text width=6em
  },
  measurebox/.style={
    meta box,
    rectangle,
    inner sep=5pt,
    minimum height=3em,
    minimum width=1.8cm,
    align=center,
    text width=1em
  },
  judgebox/.style={
    meta box,
    diamond,
    aspect=2,
    minimum height=3em,
    minimum width=2cm,
    align=center,
    text width=2.5em
  },
  round box/.style={
    meta box,
    circle
  },
  every fit/.style={
    draw,
    thick,
    dashed,
    gray,
    inner sep=10pt
  },
  base/.style args={#1/#2/#3}{%
     shape=rectangle split,
     draw, inner sep=0.5mm, outer sep=0mm,
     rectangle split parts=2,
     rectangle split part fill={red!20,white},
     rectangle split draw splits=false,
     minimum width=#1,
     text width=#1-4mm,
     rectangle split part align={left, right},
     node contents={\nodepart{one}  #2
                    \nodepart[align=right]{two}  #3}
}
}

\tikzset{
  add dimen/.code 2 args={%
    \pgfkeysgetvalue{/pgf/minimum #1}\tikz@dimen@min
    \expandafter\tikz@expand@dimen\expandafter{\tikz@dimen@min + #2 * 2em}{#1}%
  },
  wider/.style={add dimen={ width}{#1}},
  higher/.style={add dimen={ height}{#1}},
}
\makeatother

\begin{figure}[bp]
  \centering
  
\begin{tikzpicture}
  \node[stepbox, wider=3, higher=1]
    (prep) {Prepare a qubit: $\{\ket{H},\ket{V},\ket{D},\ket{R}\}$};
  \node[punkt, wider=1, below=1.5 of prep, fill=gray!20!white]
    (qm) {``Quantum memory"};
  \node[basisbox, wider=3, below=1.5 of qm]
    (bc) {Basis choice: $X, Y, Z$};
  \node[measurebox, wider=1, below=1.5 of bc]
    (meas) {};
  \draw[black,very thick,-stealth] (meas) ++(0,-1.5em) -- ++(1.3em,2.5em);
  \draw[black,very thick] (meas) ++(0.9cm,-1.em) arc (0:180:0.9 and 0.5);

  \fill [gray, opacity=0.2,rounded corners] (meas) ++(3.8cm,0.2) rectangle ++(7.1,6.5);

  \begin{scope}[xshift=6cm, yshift=-3.15cm]
    \coordinate (qmC) at (node cs:name=qm,anchor=center);
    \draw (qmC) ++ (6cm,-0.8cm) node[basisbox, wider=3]
    (evebc) {Basis choice: $X, Y, Z$};
    \node[measurebox, wider=1, right=1. of evebc]
    (evemeas) {};
    \draw[black,very thick,-stealth] (evemeas) ++(0,-1.5em) -- ++(1.3em,2.5em);
    \draw[black,very thick] (evemeas) ++(0.9cm,-1.em) arc (0:180:0.9 and 0.5);
    \draw (evemeas) ++(0,-1) node{Result $=R$};
    \coordinate (bcC) at (node cs:name=bc,anchor=center);
    \draw (bcC) ++ (6cm,0) node [judgebox] (jud) {Same?};
    \coordinate (measC) at (node cs:name=meas,anchor=center);

    \node [outbox,below=0.9 of jud] (no) {No output};
    \node[outbox, wider=1, right=-0.04 of no]
    (yes) {Output $=R$};
  \end{scope}


  \path[arrows={-Stealth[scale=1.2]}, thick]
  (prep)    edge ["Input"] (qm)
  (qm)    edge ["Output"] (bc)
  (bc)    edge (meas)
  (evebc) edge (evemeas)
  ;
  \draw[thick,arrows={-Stealth[scale=1.2]}, dashed] (bc) -- node[anchor=north]{Eavesdropping} (jud);
  \draw[thick,arrows={-Stealth[scale=1.2]}, dashed] (evebc) -- (jud);
  \draw[thick,arrows={Stealth[scale=1.2]-}] (evebc) -- ++(0,2cm) node [anchor=south]{Input};
  \draw[thick,arrows={-Stealth[scale=1.2]}, dashed] (jud) -- node[anchor=west]{No} (no);
  \draw[thick,arrows={-Stealth[scale=1.2]}, dashed] (jud) -| node[anchor=west]{Yes} (yes);
  \coordinate (resu) at (node cs:name=no,anchor=south east);
  \draw[thick,arrows={-Stealth[scale=1.2]}, dashed] (resu) ++(-0.5pt,0) |- (meas);
  \draw (meas) ++(3.5cm,-0.3) node {Faked-state attack};
  \draw[->,thick, double distance=1pt,>={Latex[length=0pt 3 0]}] (qm) ++( 1.5cm,0)-- node[anchor= north]{Inside} ++(2cm,0);

  

\end{tikzpicture}

\caption{The scheme to fake a quantum memory.}
\label{fig:attack}
\end{figure}